\documentclass{PoS}
\def\Tr{{\rm Tr}}
\title{Baryon correlators containing different diquarks from lattice
simulations}

\ShortTitle{Baryon correlators containing different diquarks from
lattice simulations}

\author{\speaker{Zhaofeng Liu} and Thomas DeGrand\\
        Department of Physics, University of Colorado, Boulder, CO 80309
USA\\
        E-mail: \email{zhaofeng.liu@colorado.edu},
        \email{degrand@pizero.colorado.edu}}

\abstract{Point to point vacuum correlators containing diquarks
in the color anti-triplet representation are
computed both in the quenched approximation and dynamical overlap
simulations with two flavors. The scalar, pseudoscalar and axial vector
diquarks are combined with light quarks to form color singlets. The
scalar (``good") diquark channel
shows a stronger attraction than the axial vector (``bad") channel in
the quenched data set. The
pseudoscalar diquark channel shows a finite volume
zero mode artifact: the correlator becomes negative at large distance
when the quark mass is small.
By separating configurations without zero modes from those
with zero modes, we found that the zero modes have an important
contribution in both the attraction in the scalar channel and the
repulsion in the pseudoscalar channel. In the
axial vector diquark channel, we did not find apparent zero mode effects.}

\FullConference{XXIVth International Symposium on Lattice Field Theory\\
                July 23-28, 2006\\
                Tucson, Arizona, USA}

\begin{document}

\section{Introduction}\label{Intr}
It is possible that bound states of two quarks, diquarks, may be a
component of the structure of hadrons. For a review of their
phenomenology, see~\cite{Anselmino:1992vg}.
Although the constituent quark model has enjoyed a rather
successful
description of a huge body of hadronic states, it cannot explain all
aspects of hadron phenomenology. One example is the paucity of exotics
which are states beyond $qqq$ baryons and $\bar{q}q$ mesons. As a
special case: why do not
a proton and a neutron in close contact merge into a single state?
Recently, the possible existence of
exotic particles~\cite{Jaffe:2003sg, Maiani:2004uc} such as the 
pentaquark~\cite{Danilov:2005kt} has
motivated  research, e.g. ~\cite{Cristoforetti:2004kj, Shifman:2005wa,
Alexandrou:2005zn, Orginos:2005vr, Fodor:2005qx}, to understand
the role of diquarks in QCD.

In this work, we study diquarks by investigating the properties of 
 point to point vacuum baryon correlators
in quenched and two flavor dynamical overlap simulations. We
calculate several current correlators in which two quarks are 
combined into 
scalar, pseudoscalar and axial vector diquarks in the color
anti-triplet representation. Each correlator is normalized by its free
field theory version so that we can see if there is attraction or
repulsion in a specific channel. For each channel in the
quenched data set, we
separate the configurations with zero modes from those without zero
modes. By comparing the correlators from the two groups, we can
try to detect zero mode effects. The dynamical simulation is done with
fixed topological charge $Q=0$ or $\pm1$. i.e., there is no zero mode or
one zero mode with positive (negative) chirality in each data set. 
By comparing the correlators from different $Q$ sectors, we can again
examine the zero mode effects.

\section{Methodology}
\label{dimeth}
Diquarks are not color singlets. To investigate their properties
on the lattice,
a diquark can be combined with a static
(infinite heavy) quark to form a colorless state as was done in
Refs.~\cite{Alexandrou:2005zn, Orginos:2005vr}. 
In Ref.~\cite{Hess:1998sd}, gauge dependent diquark
correlation functions were calculated in a fixed gauge, the Landau
gauge, on the lattice. Whatever a diquark is, the environment it
 feels in a baryon with one heavy quark is different from that in
a light baryon with three light quarks.
Here we combine a light quark with
a diquark with a specific quantum number in the color anti-triplet
representation to
get a color singlet
and compute the point to point correlation function. 
By doing this, we cannot extract the mass
splitting between spin 0 and spin 1 diquark states or the size of a
diquark state as were done in Refs.~\cite{Alexandrou:2005zn,
Orginos:2005vr, Hess:1998sd} since the interaction between a diquark and
a light quark also depends on the spin of the diquark. However, by
comparing baryon correlators containing different diquarks, we can see which
diquark is favored and which is not. We can also
investigate the effects of the zero modes. Excess zero modes
of Dirac operators are  artifacts in quenched simulations.

The currents and correlation functions we considered are collected
in Table~\ref{currenttab}. Here $C$ is the charge-conjugation operator.
\begin{table}
\caption{Currents and correlation functions}
\begin{center}
\begin{tabular}{ccll}
\hline\hline
$J^P$ (diquark) & Color & Current & Correlator $R(x)$ \\
\hline
$0^+$ & $\bar{3}$ & $J^5=\epsilon_{abc}[u^aC\gamma_5d^b]u^c$ &
$\frac{1}{4}\Tr[\langle\Omega|TJ^5(x)\bar{J}^5(0)|\Omega\rangle
x_\nu\gamma_\nu]$ \\
$0^-$ & $\bar{3}$ & $J^I=\epsilon_{abc}[u^aCd^b]u^c$ &
$\frac{1}{4}\Tr[\langle\Omega|TJ^I(x)\bar{J}^I(0)|\Omega\rangle
x_\nu\gamma_\nu]$ \\
$1^+$ & $\bar{3}$ & $J^3=\epsilon_{abc}[u^aC\gamma_3d^b]u^c$ &
$\frac{1}{4}\Tr[\langle\Omega|TJ^3(x)\bar{J}^3(0)|\Omega\rangle
x_\nu\gamma_\nu]$ \\
\hline\hline
\end{tabular}
\label{currenttab}
\end{center}
\end{table}
We choose not to consider diquarks in the color sextet representation
because they have much larger color electrostatic energy and thus are
not favored phenomenologically~\cite{Jaffe:2004ph}. 
In perturbative QCD, one-gluon exchange leads to attraction
~\cite{DeRujula:1975ge, DeGrand:1975cf} between
two quarks in the color anti-triplet representation with $J^P=0^+$.
Instanton interactions~\cite{'tHooft:1976fv, Shuryak:1981ff, Schafer:1996wv}
is also
attractive in the scalar diquark channel. Because it is thought to be
attractive, this channel is called the ``good"
diquark. We use the current $J^5$ in
Table~\ref{currenttab} to investigate this channel. In quenched
simulations~\cite{DeGrand:2001tm}, the point to point scalar meson correlator
was found to be negative, a quenching artifact. We are curious to see if we
will see similar artifact in the pseudoscalar diquark channel by using
$J^I$. (This effect is predicted by instanton liquid models.) 
The $1^+$ diquark is called the ``bad" diquark in the literature, since
all models suggest that it is heavier than the scalar diquark.

The two point correlator for a current $J$ is defined as
$\langle\Omega|TJ(x)\bar{J}(0)|\Omega\rangle$,
where $|\Omega\rangle$ is the vacuum and $T$ is the time order operator.
For free massless quarks, the quark propagator in coordinate space takes
the form (in Euclidean space)
\begin{equation}
\langle0|Tq(x)\bar{q}(0)|0\rangle=\frac{1}{2\pi^2}\frac{x_\mu\gamma_\mu}{x^4}.
\end{equation}
Here $\mu$ is summed over.
Then for the current $J^5$ in Table~\ref{currenttab}, we have
\begin{equation}
\langle0|TJ^5(x)\bar{J}^5(0)|0\rangle=-\frac{15}{4\pi^6}
\frac{x_\mu\gamma_\mu}{x^{10}}.
\label{j5free}
\end{equation}
Similarly, we can get the free correlators for other currents in
Table~\ref{currenttab}. They are the same as the result in
Eq.(\ref{j5free}) except for a sign flip for $J^I$ 
(for $J^3$, $x_3=0$ is needed to get a same result). As was done in
Ref.~\cite{Chu:1993cn, Schafer:1993ra}, 
it is convenient to multiply the correlators
with
$x_\nu\gamma_\nu$ and take the trace in the Dirac indices to get
a number for each of the currents. For example, from Eq.(\ref{j5free}),
we find
$R_0(x)\equiv\frac{1}{4}\Tr[\langle0|TJ^5(x)\bar{J}^5(0)|0\rangle
x_\nu\gamma_\nu]=-15/4\pi^6x^8$.
$R_0(x)$ is used to normalize the
interacting correlator $R(x)$, i.e. we will examine the ratio
$R(x)/R_0(x)$ for each current.
In our lattice simulations, we use the free lattice correlators $R_0(x)$ to
do the normalization to reduce lattice artifacts.

The point source for quark propagators is put on $(0,0,0,0)$.
The boundary condition in the time direction is anti-periodic, while 
in the space direction it is periodic. To avoid different boundary
condition effects, we fix
the time component $n_t$ of $x$ in $R(x)$ at zero,
i.e. $x=(n_x,n_y,n_z,0)$. We take all the diagonal points $(n,n,n,0)$
and those lying approximately along the diagonal: $(n,n-1,n-1,0)$ and
$(n,n,n-1,0)$.

Our quenched data set has 40 configurations with lattice size $16^4$
and gauge coupling $\beta=6.1$. The bare quark mass $am_q=0.015$, 0.025
and 0.05. The lattice spacing is 0.08 fm 
determined from the Sommer parameter.
The dynamical data set has about 30
configurations
for each bare quark mass $am_q=0.015$, 0.03 or 0.05 and for each topological
charge sector $|Q|=0$ or 1. The lattice size is $10^4$ and $\beta=7.2$.
The lattice spacing is rather coarse: $a\sim 0.16$ fm.

\section{Results and discussion}
\label{results}
\begin{figure}
{\centering\includegraphics[width=72mm,height=60mm]
{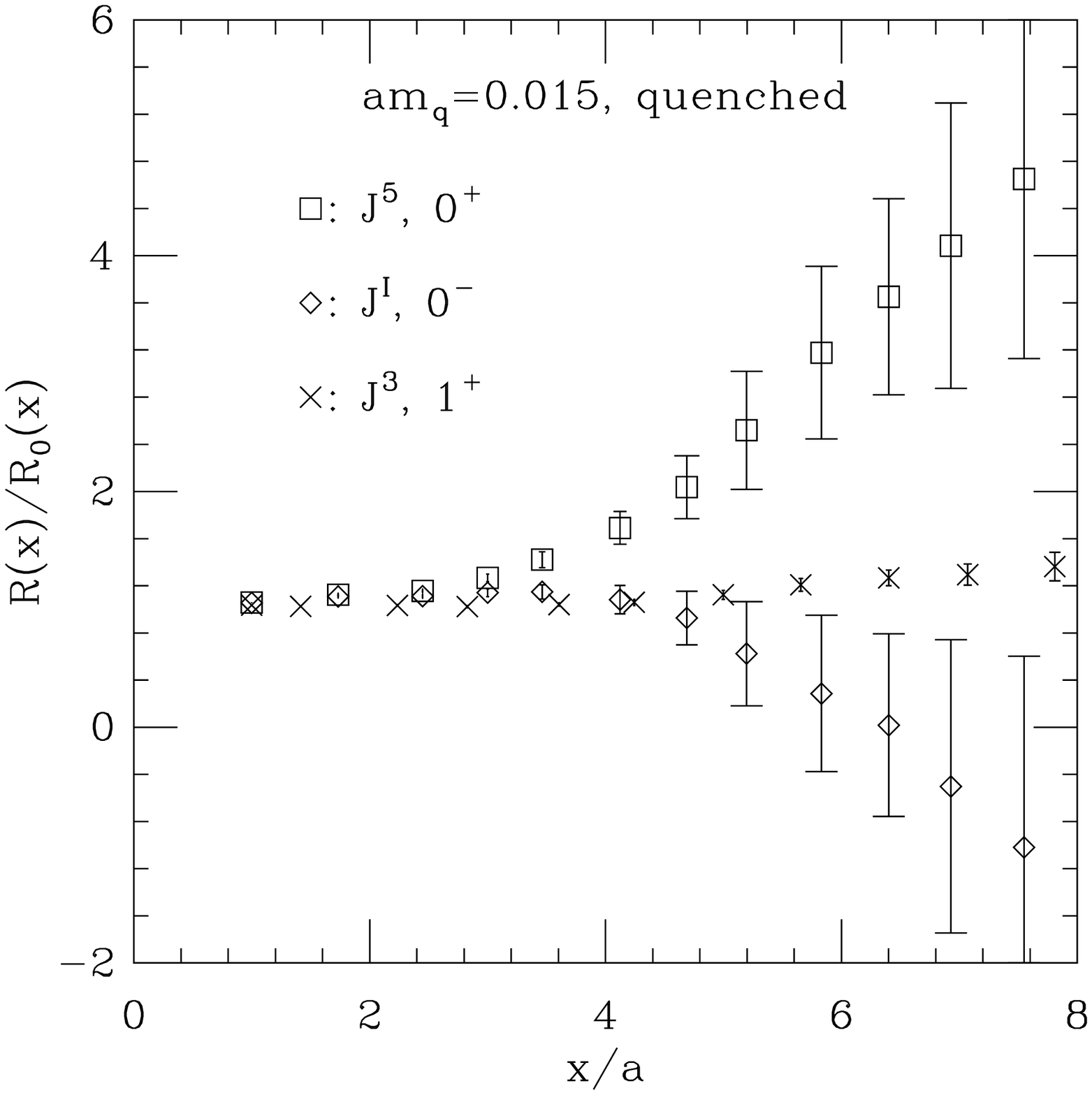}
\includegraphics[width=72mm,height=60mm]
{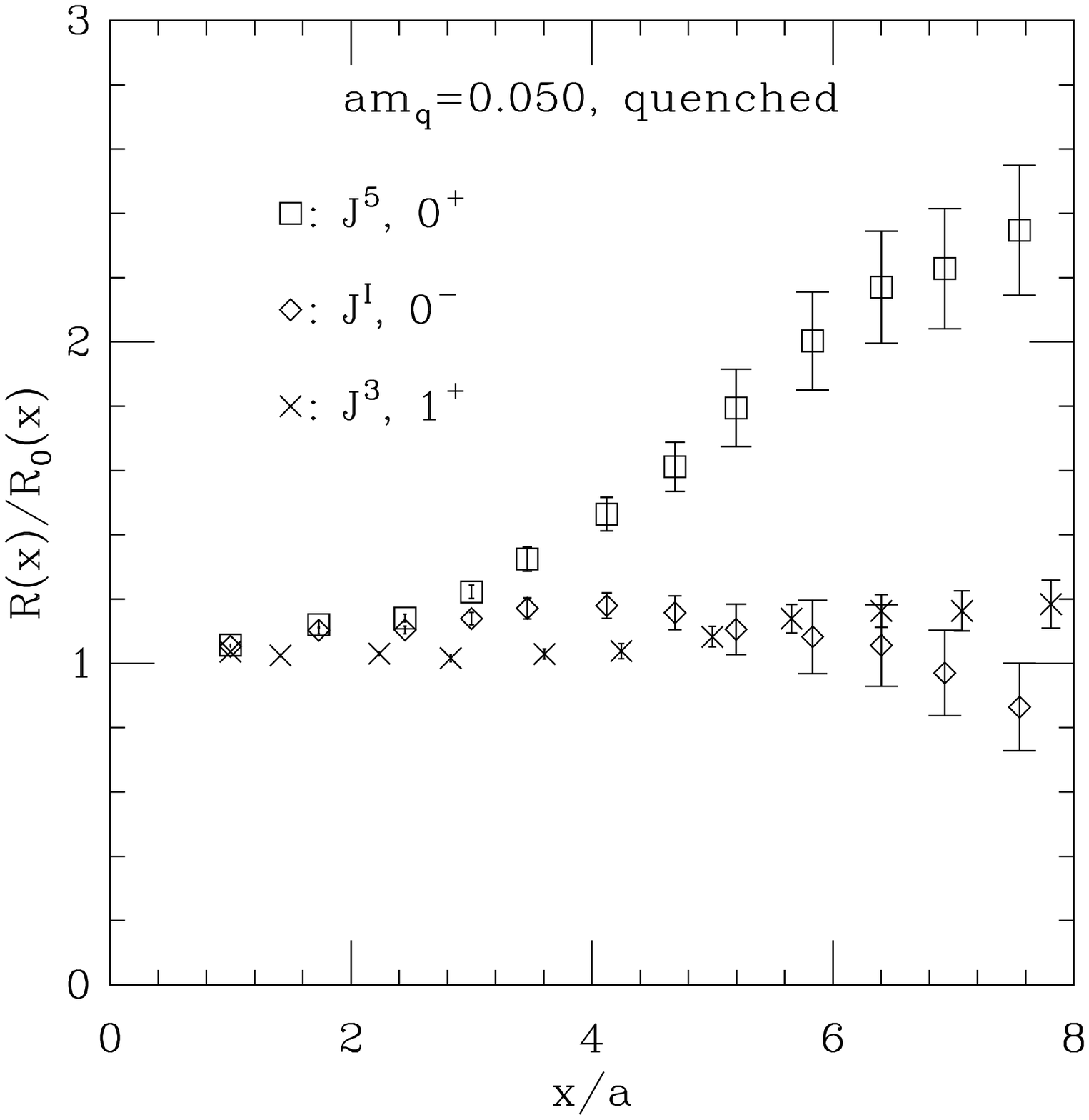}}
\caption{Comparison of the normalized correlation functions of the
three currents from the quenched data set. 
The squares are for $J^5$ which contains
a scalar (``good") diquark structure. The diamonds are for 
$J^I$ which contains a pseudoscalar diquark structure.
The crosses are for $J^3$ which contains an axial vector (``bad") diquark
structure.}
\label{compall3}
\end{figure}

Fig.~\ref{compall3} shows the normalized correlation functions of the
three currents from the quenched data set for quark mass $am_q=0.015$
and 0.05. As we can see, there is a strong attraction
in the scalar diquark channel, which is similar to the attraction seen
in the pseudoscalar meson
channel in Refs.~\cite{Chu:1993cn, Hands:1994cj, DeGrand:2001tm}. Also
we can see a quark mass dependence in this channel. As the quark
mass decreases, the attraction increases.
Since our lattice spacing $a$ is 0.08 fm, the region we are looking at
is 0 fm $<x<$ 0.64 fm. The
correlator for $J^I$ has a tendency to go negative at 
large distance as the quark mass decreases. This behavior is very
similar to what observed for the correlator of the scalar meson in
Ref.~\cite{DeGrand:2001tm}. It was argued in Ref.~\cite{DeGrand:2001tm}
that zero modes are the source of the negativity. We will investigate the
zero mode contributions in all three channels later. The correlator for
$J^3$ is flat and stays close to one. This means that there is little
correlation among quarks in this channel.

\begin{figure}
{\centering\includegraphics[width=72mm,height=60mm]
{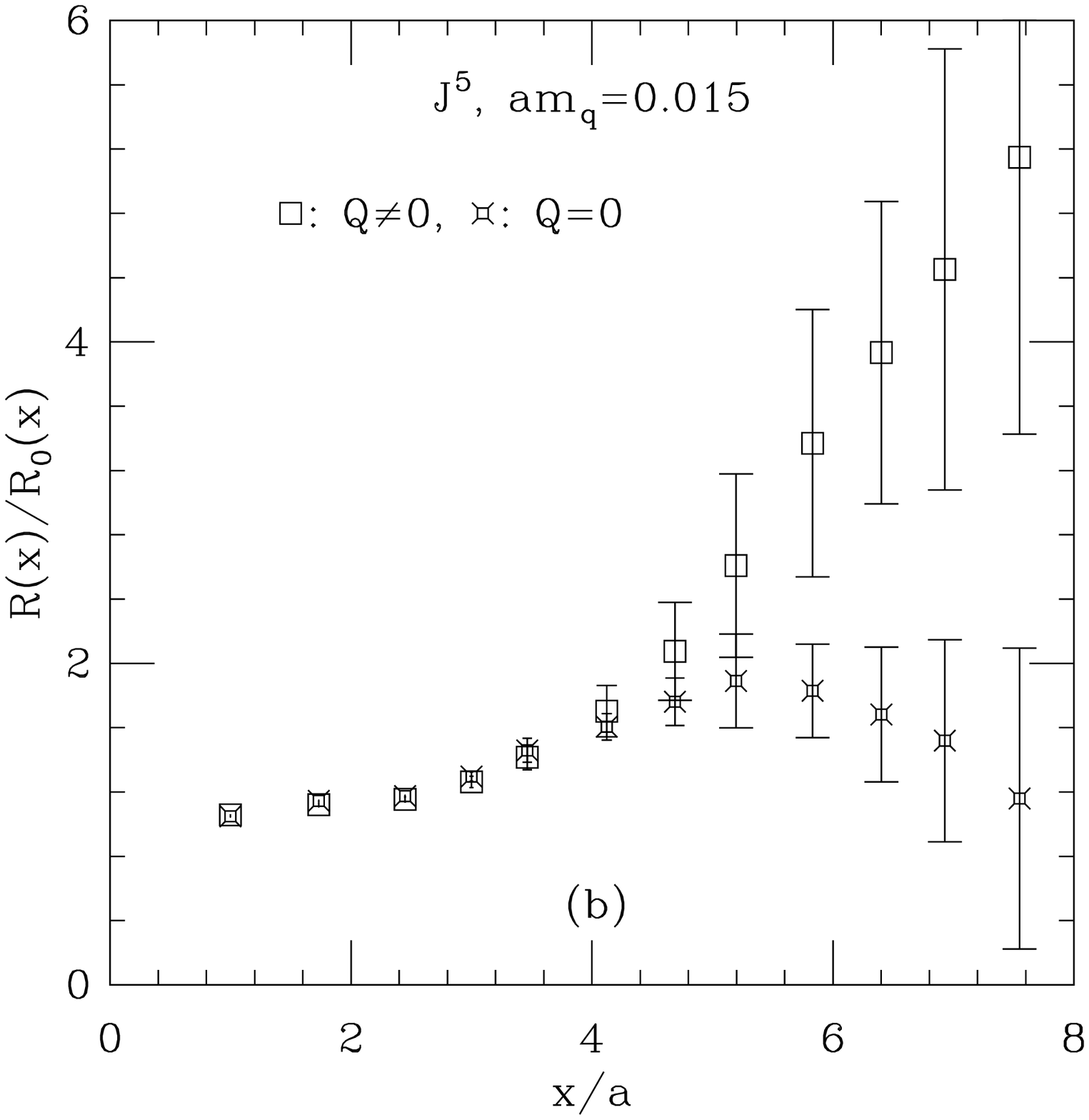}
\includegraphics[width=72mm,height=60mm]
{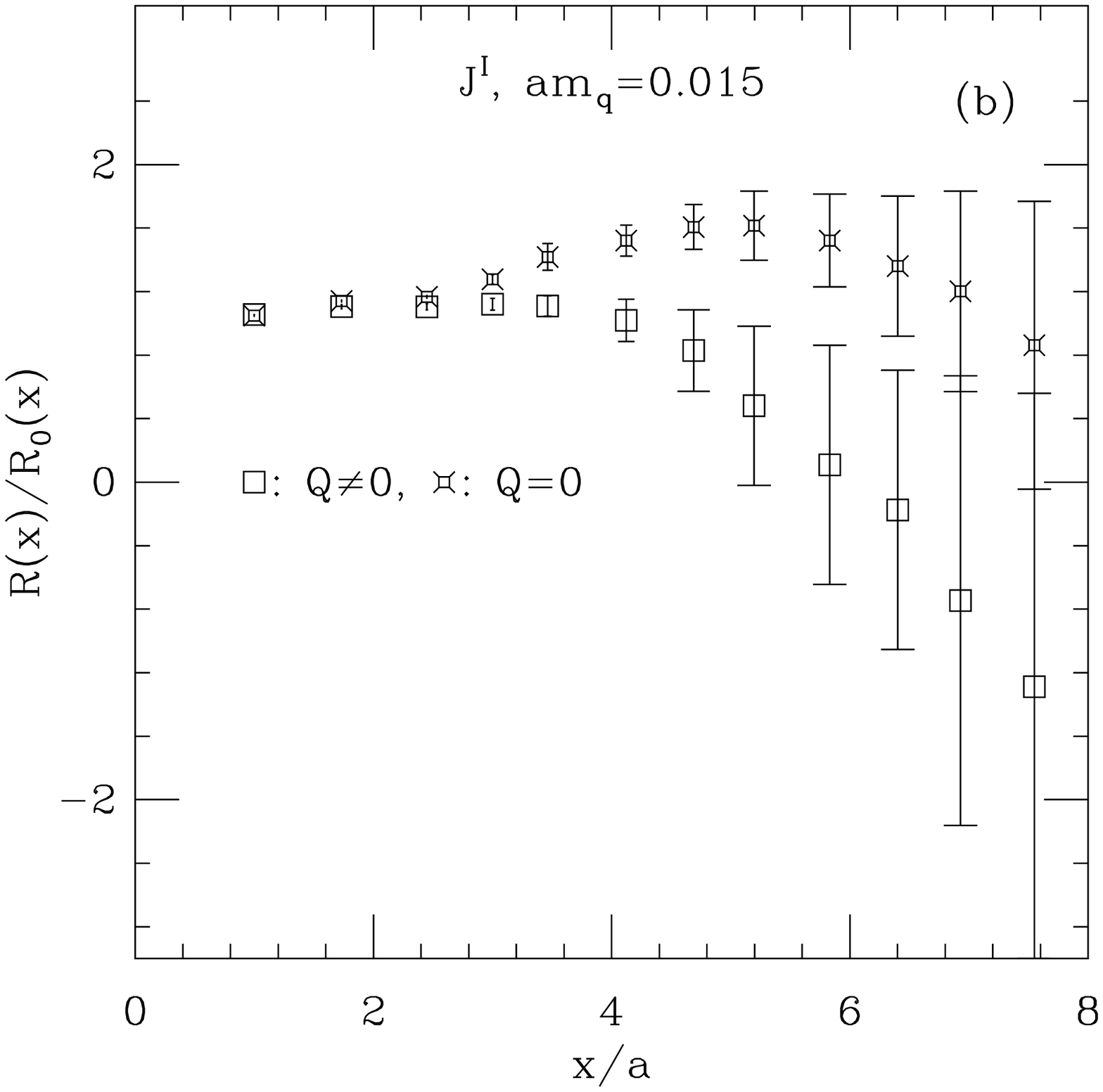}}
\caption{Comparison of correlators from the topological charge
$Q\neq0$ sector (squares) and the $Q=0$ sector (fancy squares)
in the quenched data set.
The graph on the left is for $J^5$, the one on the right is for
$J^I$.}
\label{zeromodes}
\end{figure}

\begin{figure}
\begin{center}
\includegraphics[width=72mm,height=60mm]
{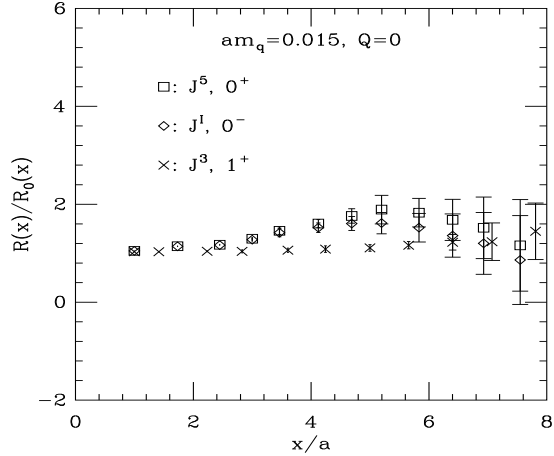}
\caption{Comparison of the correlators of the three currents from the topological charge
$Q=0$ sector in the quenched data set. The three channels are not so different
when there is no zero mode contribution. It seems that there is more attraction in channel
$J^5$ and $J^I$ than in channel $J^3$ around $x\sim4a=0.32$ fm.}
\end{center}
\label{Q0comp}
\end{figure}

To see zero mode effects, we separate the 40 configurations in the
quenched data set into two
groups: 35 of them with topological charge $Q\neq0$ and the other 5 with
$Q=0$. i.e. the first group contains zero modes while the second one
does not. From each group we compute the correlation functions. The
results are compared in Fig.~\ref{zeromodes} for quark mass $am_q=0.015$.
The attraction in the current $J^5$ at large $x$ 
is mainly from zero mode contributions. The repulsion in the current
$J^I$ also has a big contribution from zero modes. At quark mass
$am_q=0.05$, the correlators from the two groups agree with each other
within error bars (no graph is shown here).
It is not surprising that zero mode effects become large at small quark
masses since the zero mode
contribution to the quark propagator is proportional to the inverse 
of the quark mass. For the current $J^3$, we do not see apparent zero
mode effects.
In Fig.~\ref{Q0comp} we compare the three correlators obtained
from configurations without zero modes. They are not so different
as in Fig.~\ref{compall3} when there is no zero mode contribution.
Nevertheless, it seems that there is more attraction in channel
$J^5$ and $J^I$ than in channel $J^3$ around $x\sim4a=0.32$ fm.

The results from the two flavor dynamical simulation cannot be compared
to the quenched results directly because the dynamical simulation is 
done with fixed
topological charge $|Q|=0$, or 1. Thus we will only compare the
dynamical results from different $Q$ sectors to investigate the zero mode
effects.

\begin{figure}
{\centering\includegraphics[width=72mm,height=60mm]
{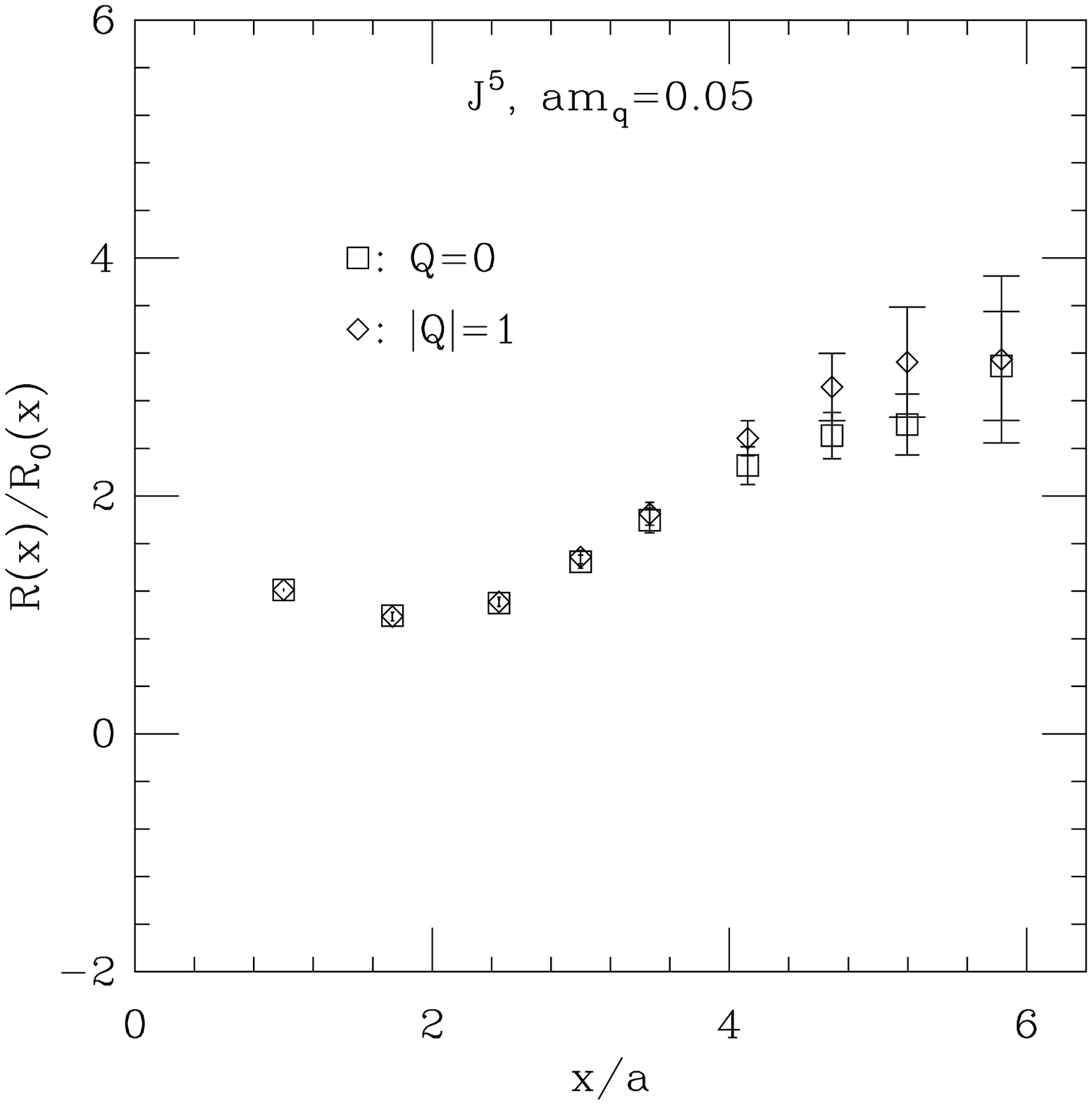}
\includegraphics[width=72mm,height=60mm]
{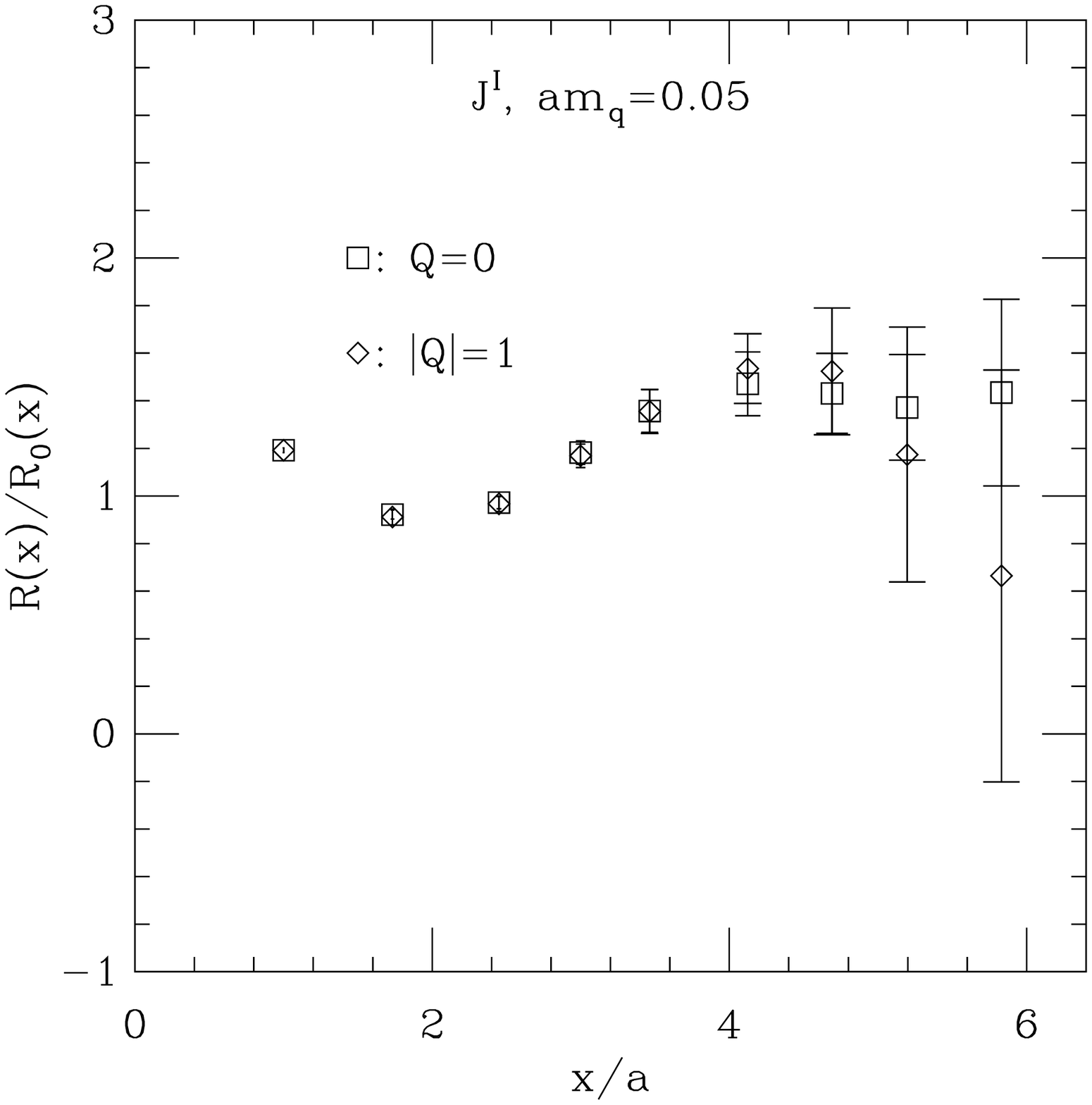}}
\caption{Comparison of correlators from the topological charge
$Q=0$ sector (squares) and the $|Q|=1$ sector (diamonds) from
two flavor dynamical simulations.
The graph on the left is for $J^5$, the one on the right is for
$J^I$. There is no difference within error
bars between the correlators from the two sectors.}
\label{dyn2flavor}
\end{figure}

\begin{figure}
{\centering\includegraphics[width=72mm,height=60mm]
{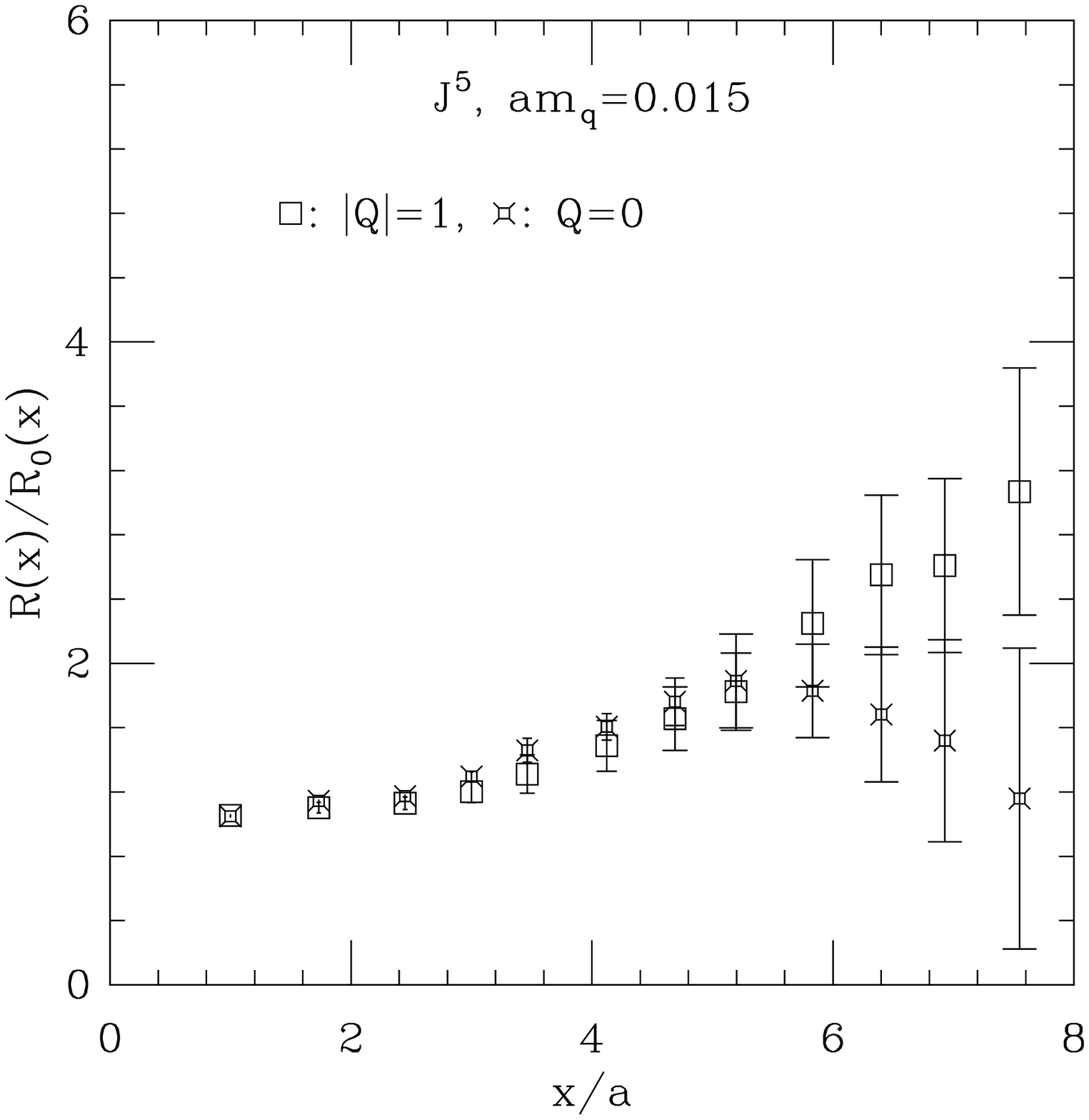}
\includegraphics[width=72mm,height=60mm]
{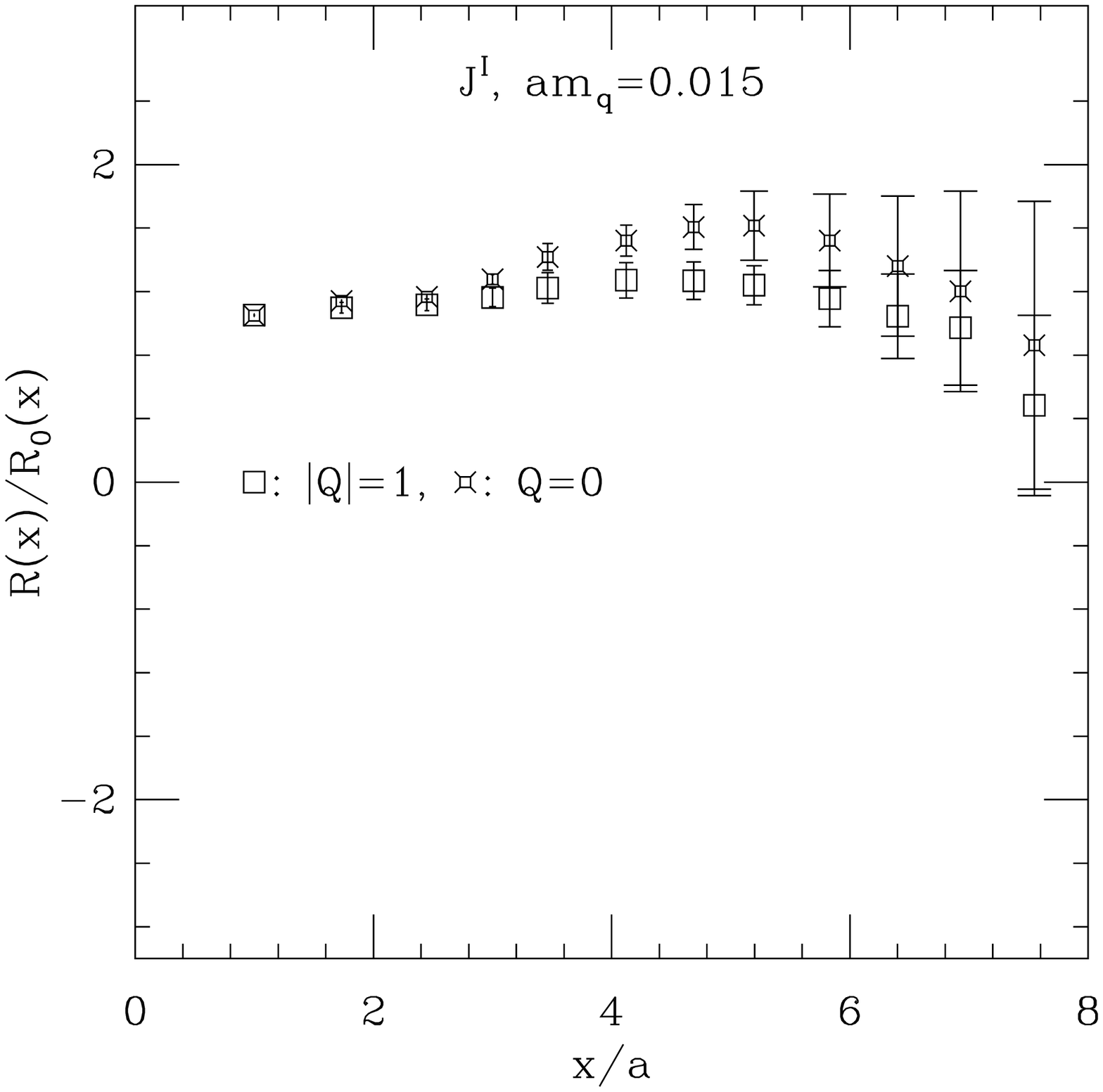}}
\caption{Comparison of correlators from the topological charge
$Q=0$ sector (fancy squares) and the $|Q|=1$ sector (squares) from
the quenched data set.
The graph on the left is for $J^5$, the one on the right is for
$J^I$. The difference between the correlators from the two topological
sectors is small.}
\label{Q01quenched}
\end{figure}

The comparisons of the correlators from the two topological sectors
are given in Fig.~\ref{dyn2flavor}. There is no difference within error
bars between the correlators from the two sectors for both current $J^5$
and $J^I$. Similarly, no difference is seen for the current $J^3$.
In Fig.~\ref{zeromodes}, the $Q\neq0$ sector of the quenched
simulation contains not only
configurations with $|Q|=1$ but also those with $|Q|>1$. To compare with
the dynamical simulation results here, we pick out the $|Q|=1$ and $Q=0$
configurations and calculate the correlators again. The results are
given in Fig.~\ref{Q01quenched}.
The difference between the two sectors is small,  just like the
dynamical results in Fig.~\ref{dyn2flavor}. Comparing Fig.~\ref{Q01quenched} with 
Fig.~\ref{zeromodes}, we see that the attraction in the $J^5$
channel and the repulsion in the $J^I$ channel in the quenched simulation are
mainly from configurations with big topological charge: $|Q|>1$.

\section{Conclusions}
The quenched simulation shows that
the attraction in the $J^5$ channel
is the strongest. $J^5$ contains a scalar diquark structure.
By comparing the correlators from different topological sectors,
we found that
both the attraction in the scalar diquark channel and the
repulsion in the pseudoscalar diquark channel have big
contributions from configurations with more than one zero mode.
The correlators for the currents $J^5$ and $J^I$ obtained from configurations 
without zero modes in the quenched data set 
are similar to each other. Both of them seem to
have more attraction around $x\sim4a=0.32$ fm than the correlator
for the current $J^3$ which contains an axial vector diquark structure.
The correlators from the two flavor dynamical data set show no difference
between the $Q=0$ and $|Q|=1$ sector. This confirms that zero mode effects
are mainly from configurations with big topological charges. In this respect,
quenched and full QCD in sectors of low $Q$ are not too different.

In full QCD, small quark masses suppress high $Q$ configurations. Since they are
absent there, we suspect that different diquarks in light baryons are not
that different.
At a minimum, diquark contributions in quenched QCD ({\it not} filtered by $Q$)
and full QCD are different,
and results from quenched simulations may be misleading.

\section*{Acknowledgments}
This work was supported by the US Department of Energy.

\end{document}